\documentclass[iop,apj]{emulateapj}
\usepackage{amsmath,amssymb,amstext}
\usepackage[breaklinks,colorlinks,citecolor=blue,linkcolor=magenta]{hyperref} 

\usepackage[all]{hypcap}
\def\baselinestretch{0.96}
\usepackage{graphicx}
\usepackage{subfigure}

\usepackage{float}

\begin{document}

\title{First Results on the Epoch of Reionization from First Light with SARAS 2}
\author{Saurabh Singh$^{1,\dag}$}
\author{Ravi Subrahmanyan$^1$}
\author{N. Udaya Shankar$^1$}
\author{Mayuri Sathyanarayana Rao$^1$}
\author{Anastasia Fialkov$^2$}
\author{Aviad Cohen$^3$}
\author{Rennan Barkana$^3$}
\author{B.S. Girish$^1$}
\author{A. Raghunathan$^1$}
\author{R. Somashekar$^1$}
\author{K.S. Srivani$^1$}
\thanks{$^\dag$Joint Astronomy Program, Indian Institute of Science, Bangalore 560012, India }
\affil{{\small $^{1}$Raman Research Institute, C V Raman Avenue, Sadashivanagar, Bangalore 560080, India}}
\affil{{\small $^{2}$Harvard-Smithsonian Center for Astrophysics, Institute for Theory and Computation, 60 Garden Street, Cambridge, MA 02138, USA}}
\affil{{\small $^{3}$Raymond and Beverly Sackler School of Physics and Astronomy, Tel Aviv University, Tel Aviv 69978, Israel}}
\email{Email of corresponding author: saurabhs@rri.res.in}

\begin{abstract}

Long wavelength spectral distortions in the Cosmic Microwave
Background arising from the 21-cm transition in neutral Hydrogen are a
key probe of Cosmic Dawn and the Epoch of Reionization. These features
may reveal the nature of the first stars and ultra-faint galaxies that
transformed the spin temperature and ionization state of the
primordial gas. SARAS~2 is a spectral radiometer purposely designed
for precision measurement of these monopole or all-sky global 21-cm
spectral distortions. We use 63~hr night time observing of the radio
background in the frequency band 110-200~MHz with the radiometer
deployed at the Timbaktu Collective in Southern India to derive
likelihoods for plausible redshifted 21-cm signals predicted by
theoretical models. First light with SARAS 2 disfavors the class of models that
feature weak X-ray heating (with $f_X \leq 0.1$) and rapid reionization (with peak $\frac{dT_b}{dz} \geq 120~\textrm{mK per unit redshift interval}$    ). 

\end{abstract}


\section{Introduction}

The Epoch of Reionization (EoR), beginning with first light from the
first stars and ultra-faint galaxies, and ending with almost complete
reionization of the primordial gas, marks an important period in the
cosmic evolution of baryons
\cite[]{2001PhR...349..125B,2005SSRv..116..625C,2013ASSL..396...45Z,2016ARA&A..54..313M,Haiman2016,
  2016PhR...645....1B}. There is considerable uncertainty and limited
observational constraints on the astrophysical evolution in this
period, including the nature of the first sources of light, and the
thermal and ionization state of the intergalactic medium (IGM).

Current observational constraints on the EoR are either indirect or
integrated in time. They include the Gunn-Peterson trough towards
high-redshift QSOs \cite[]{2006ARA&A..44..415F,2015MNRAS.447..499M},
which places the end of reionization at redshift $z \sim 6$; the
evolution in the luminosity function of Lyman-$\alpha$ galaxies, which
indicates an ionization fraction of 0.4--0.6 at $z \sim 7$
\cite[]{2017arXiv170302985Z}; detection of the EoR signature in the
Cosmic Microwave Background anisotropies, placing the average redshift of
reionization $z_r$ between 7.8 and 8.8
\cite[]{2016A&A...596A.108P,0067-0049-208-2-19}; and upper limits on
the kinematic Sunyaev-Zeldovich effect, limiting the extent of the EoR
to $\Delta z_r < 2.8 $ \cite[]{2016A&A...596A.108P}.

On the other hand, the redshifted 21-cm line from neutral Hydrogen is
a direct probe of the state of the gas in the EoR.  Wouthuysen-Field
\cite[]{Wouthuysen:1952,Field:1958} coupling of the spin to the
kinetic temperature via Lyman-$\alpha$ photons, gas heating via X-rays
and reionization via ultra-violet radiation generate spatial and
temporal fluctuations in the 21-cm signal, all of which result in a
redshifted 21-cm power spectrum whose monopole or all-sky global
component traces the mean cosmological evolution \cite[]{MMR}.
Although a direct detection of the 21-cm signal continues to be
elusive, the PAPER radio interferometer derived lower limits in the
range 5--10~$\textrm{K}$ on the IGM temperature at $z=8.4$ based on
upper limits to the power spectrum of 21-cm spatial fluctuations at
that epoch; the derived limit depends on the assumed ionization
fraction \cite[]{2015ApJ...809...62P}.  Recently, additional upper
limits on the 21-cm power spectrum have been reported by MWA
\cite[]{2016ApJ...833..102B} and LOFAR \cite[]{2017arXiv170208679P}.

While these experiments as well as HERA and SKA-Low work towards
detection of the 21-cm power spectrum, detection of the global 21-cm
signal from the EoR could well prepare the way with useful constraints
on the mean evolution
\cite[]{2006PhR...433..181F,2006MNRAS.371..867F,2012RPPh...75h6901P},
given that a single-element radiometer suffices for a
detection and the time required for achieving useful sensitivity,
under ideal conditions, is only a few minutes
\cite[]{1999A&A...345..380S,2017ApJ...840...33S}. The present
uncertainty in the astrophysical parameters during the EoR allows for
various possible global 21-cm signals. Hence, well-calibrated wide-band radiometer measurements could pin
down the underlying astrophysics \cite[]{2012MNRAS.424.2551M,
  2013ApJ...777..118M, 2016MNRAS.455.3829H, 2016arXiv160902312C}.

A non-detection by the EDGES experiment, which targeted the global
signal, placed a lower limit on the extent of reionization of $\Delta
z_r > 0.06$ \cite[]{2010Natur.468..796B}. Analysis of such
measurements up to $z \sim $12--15 used to rely on theoretical
predictions \cite[]{MMR,2006MNRAS.371..867F} that reionization
occurred in the ``saturated heating'' limit, in which cosmic heating
had occurred earlier and the IGM temperature no longer affected the
21-cm signal. However, \citet{2014Natur.506..197F} showed that late
heating, in which reionization features strong 21-cm absorption due to
a still-cold IGM, is quite plausible, opening up a wide variety of
possible 21-cm signals. Extrapolations of recent observations to high
redshift also support such scenarios
\cite[]{2016arXiv160607887M,2017MNRAS.464.1365M}.

A number of experiments are underway to detect the global 21-cm
signal, including EDGES~2 \cite[]{2017ApJ...835...49M}, LEDA
\cite[]{2016MNRAS.461.2847B}, BIGHORNS \cite[]{2015PASA...32....4S},
and SCI-HI \cite[]{2014ApJ...782L...9V}. Attaining the necessary
sensitivity to plausible signals is a formidable challenge: the
cosmological signal needs to be discerned in the presence of Radio
Frequency Interference (RFI), instrumental systematics
\cite[]{2013PhRvD..87d3002L}, ionospheric effects \cite[]{2014MNRAS.437.1056V,2015ApJ...813...18S} and Galactic and Extra-Galactic
foregrounds, which can be 5--6 orders of magnitude brighter than the
signal \cite[]{2015MNRAS.449L..21H, 2012MNRAS.419.3491L}.
Fortunately, the foregrounds have been shown to be spectrally smooth
to $\textrm{mK}$ levels in the frequency range of 40--200~MHz and can
be modeled by smooth functions
\cite[]{2017ApJ...840...33S}. Similarly, the ionospheric effects---absorption, emission, refraction and the stochastic error due to temporal variations in total electron content---result in spectrally smooth components  \cite[]{2014MNRAS.437.1056V,2015ApJ...813...18S} that may be subsumed by a smooth modeling of the foreground.  However, the level of systematics
is critically dependent on the radiometer design, calibration
scheme, as well as data modeling strategies
\cite[]{2013ExA....36..319P, 2017ApJ...835...49M,
  2015ApJ...799...90B}.

SARAS 2 is a spectral radiometer that aims to detect the global redshifted 21-cm
signal from the EoR over 40-230~MHz. Below, we describe its design philosophy,
calibration methodology, algorithms developed for RFI excision, and
modeling of the foregrounds and instrumental systematics.  We present
results from first light upon deploying the system at a relatively
radio quiet site at the Timbaktu Collective in Southern India.

\section{SARAS 2 spectral radiometer}

SARAS~2 has a wide-band wide-field monopole antenna deployed on open
level ground with receiver electronics enclosed in a unit below the
antenna and below ground.  The receiver is a correlation
spectrometer in that the antenna signal is first split into two, then
amplified separately in two parallel signal paths.  The analog signals
are transmitted on optical fiber to a signal processing unit located
100~m away, which is followed by a digital spectrometer that
spectrally decomposes the signals, computes the complex
cross-correlation between the signals and records the spectra.  The
entire system operates on batteries and can be deployed at remote
radio-quiet sites.

\subsection{The antenna}

\begin{figure}[htbp]
\begin{center}
\includegraphics[scale=0.36]{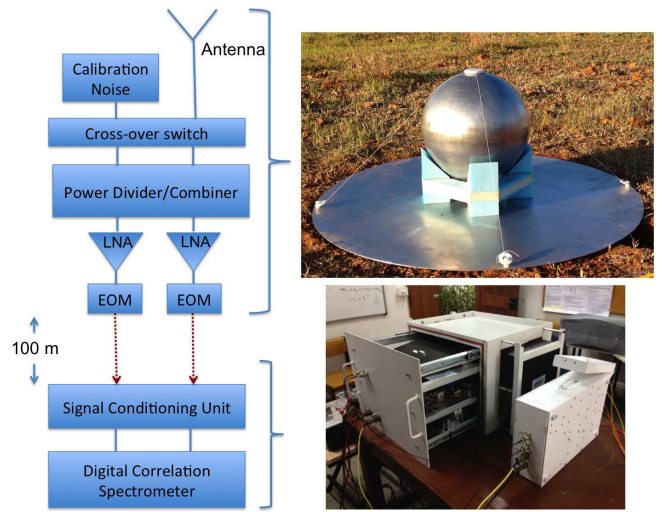}
\caption{SARAS~2: In the schematic, LNA refers to Low-Noise Amplifiers while EOM are Electro-Optical Modulators. The upper right image shows the sphere-disk monopole, with the sphere
  supported using styrofoam, cotton
  strings and teflon fasteners. The lower right image shows the spectrometer.}
\label{fig:Ant1}
\end{center}
\end{figure}

The SARAS~2 antenna is a sphere-disk monopole antenna (see
Fig.~\ref{fig:Ant1}) in which a circular aluminum disk on the ground
is one element and a sphere atop an inverted cone forms the second
element; the sphere and cone are smoothly conjoined and the cone
surface meets the sphere tangentially.  The edge of a small
circular hole at the center of the disk continues down as the outer
conductor of a coaxial cable, whose central conductor connects to the
apex of the inverted cone.  The antenna smoothly transforms into an
unbalanced transmission line that connects to the receiver below, thus
avoiding any balun or impedance transformer that could introduce frequency-dependent resistive losses, which would be difficult to characterize to the required accuracy.

The antenna is electrically small with its spherical radiating element, of diameter 0.292~m, less than $\lambda/4$ at the highest frequency. Further, the disk radius is 0.435~m, guaranteeing that internal reflection of currents from the edge can only create sinusoids of period about 350~MHz in the frequency-domain characteristics. The electrically small dimensions ensure that the entire observing band is within the first resonance, which is at 260~MHz. The structure is of simplistic design, defined by a minimal number of parameters, with smooth characteristics.

The antenna beam is omnidirectional,
with nulls towards the horizon and zenith, with a peak at
$30{\degr}$ elevation and half power beam width of $45{\degr}$.
Frequency independence of the beam is critical for this experiment in
order to avoid coupling of sky structure into spectral features. The electrically small dimensions ensure frequency independence for the antenna beam, and we have confirmed this property by range measurements and electromagnetic simulations.

A radiation efficiency $\eta_r(\nu)$ defines the frequency-dependent
coupling of the beam-weighted sky temperature $T_{\rm sky}(\nu)$ to
the antenna.  Owing to impedance mismatch between the antenna and
transmission line, only a fraction of this power---defined by a
reflection efficiency $\eta_c(\nu)$---arrives at the receiver.  The
total efficiency $\eta_t = \eta_r \times \eta_c$ determines the
received antenna temperature:
\begin{equation}
T_{a} (\nu) =  \eta_r (\nu) \eta_c (\nu) T_{\rm sky} (\nu). \label{eq:eff}
\end{equation}
Internal receiver noise appears as an additive contaminant in measured
spectra, and internal reflections of the receiver noise at the antenna
terminals result in spectral shapes for this contaminant, with the
shape dependent on the antenna reflection coefficient $\Gamma_c
(\nu)$, which is related to $\eta_c (\nu)$ as:
\begin{equation}
\eta_c(\nu) = 1 - |\Gamma_c(\nu)|^2.
\end{equation}
Thus, if $\Gamma_c$ has any low-level embedded ripples then 
both foregrounds and receiver noise contributions in measured spectra 
would have non-smooth structure. Therefore, 
critical to detection of the EoR global signal is designing $\Gamma_c$
to be spectrally smooth.  Mathematically, we require $\Gamma_c$ to be
{\it Maximally Smooth} \cite[]{2015ApJ...810....3S}.  As discussed above, 
the shape and dimensions of the antenna are chosen to make its characteristics, including $\Gamma_c$, smooth.
To the accuracy limits of the field measurements, $\Gamma_c$ is spectrally
smooth at levels better than 1 part in $10^4$, ensuring that non-smooth features in the instrument response
to the receiver noise, if any, are below the sensitivity of the observations presented here.

\begin{figure}[htbp]
\begin{center}
\includegraphics[scale=0.6]{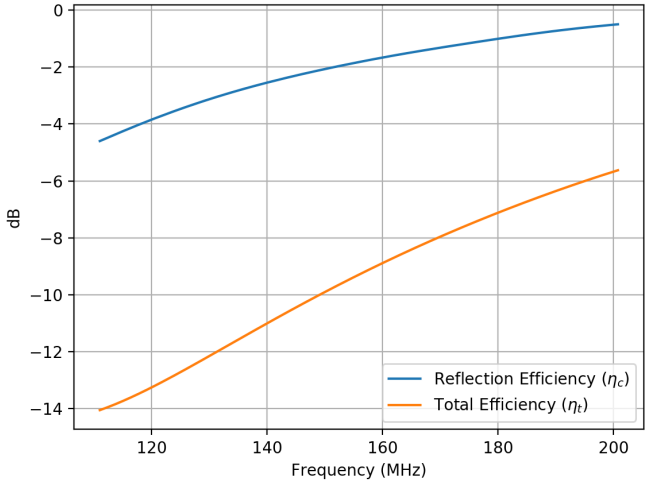}
\caption{SARAS~2 antenna efficiencies versus frequency.}
\label{fig:Ant2}
\end{center}
\end{figure} 
The total efficiency $\eta_t$ is estimated from a comparison of the
differential antenna temperature measured as the sky drifts overhead
and the expectation for this differential based on the GMOSS model for
the radio sky \cite[]{2017AJ....153...26S}. This total efficiency and
also the reflection efficiency are shown in Fig.~\ref{fig:Ant2}; the
total efficiency represents the attenuation with which any EoR
signature would be present in observed spectra.  It may be noted here
that the efficiency is poor and more so at lower frequencies; this was
a design compromise made for SARAS~2 in that efficiency was sacrificed
for spectral smoothness in the reflection efficiency and frequency
independence of the beam.

\subsection{The receiver}

The antenna signal is split coherently into two parallel paths, which
are amplified separately. The splitter also adds coherent calibration
noise into both paths; however, the relative phases of the sky and
calibration signals in the two paths differ by $180{\degr}$.  As a
consequence, the correlation spectrometer provides a difference
measurement between the sky and calibration noise
temperatures.

The configuration of the receiver is shown in Fig.~\ref{fig:Ant1}. A
cross-over switch swaps the sky and calibration signals entering the
splitter.  Differencing spectra recorded in the two positions of the
switch cancels common-mode signals entering the two parallel signal
paths.  In each position of the cross-over switch, the calibration
noise is switched on and off to provide data for bandpass calibration.
System noise couples across the parallel paths via internal
reflections at the antenna and components in the receiver chains to
give an additive spurious component in the measurement.  This is
shaped by the frequency dependence of $\Gamma_c$ and the relative path
delay with which the direct and reflected signals arrive at the
correlation spectrometer \cite[]{meys1978wave}. 

The SARAS~2 receiver
is compact, mounted directly beneath and at the antenna terminals, and
the components are interconnected thus avoiding transmission lines
in-between. The amplified signals directly modulate lasers and
transition to fibers, thus providing excellent optical isolation to
the subsequent electronics located 100~m away.  All of this helps
ensure that the additive spurious component from internal reflections and
multi-path propagation of system noise is spectrally smooth;
therefore, this unwanted component can also be modeled as a Maximally
Smooth Function.

\subsection{The digital spectrometer}

Located 100~m away from the antenna is a signal conditioning unit,
which converts the signal back to electrical from optical and limits the
band to 40--250~MHz.  This is followed by a well-shielded digital
spectrometer, which samples the pair of signals in the parallel paths
with 10-bit precision, computes 8192-point Discrete Fourier
Transforms, and measures the complex cross-correlation in each of 4096
frequency channels over the range 0--250~MHz. The signals are windowed in time domain using a Blackman-Nuttall window \cite[]{1981ITASS..29...84N}, which has been measured to suppress leakage of any RFI into the rest of the band by a factor of $10^8$.

\section{A measurement for the 21-cm EoR global signal} \label{sec:model}

SARAS~2 was deployed at the Timbaktu Collective (Latitude=+14\fdg242328,
Longitude=77\fdg612606E).  
Data were acquired over 13 nights from 2016 October to 2017 June.  Ionospheric Total Electron Content (TEC) for the entire observing was less than 20 units, corresponding to quiet conditions\footnote{CODE data archieve (ftp://ftp.unibe.ch/aiub/CODE/2016/)}.
Pre-processing and data calibration was
performed within the MIRIAD environment \cite[]{1995ASPC...77..433S} using custom tools.

Data were acquired cycling through each of four states: alternating
the cross-over switch and toggling the calibration noise in each
switch position.  A batch of sixteen 67.1-ms integrated spectra were acquired in each state of the receiver.  They were Hampel filtered \cite[]{10.2307/2285666} to reject strong RFI and then averaged. Common-mode responses of the
correlation spectrometer were rejected by differencing spectra corresponding to the
two switch states; this was followed by complex bandpass calibration.  

The calibrated spectra were processed using algorithms for
detection/rejection of data corrupted by lower levels of RFI.  Spectra
were fit with suitably high-order (10'th order) Legendre polynomials
over multiple overlapping bands, in order to fit out plausible models
for the EoR spectrum as well as foregrounds and instrumental
systematics, and outliers in the residuals were detected using median
filters and rejected. This was performed in successive iterations
while progressively lowering the detection threshold and repeating the
fits.  Data were also progressively averaged in frequency and time to
detect faint RFI that may be present in contiguous channels and/or
times. The algorithm was designed to avoid asymmetric clipping of
noise peaks that may result in bias in averaged residuals at levels at
which the EoR signal is expected.  
Rejection of data corrupted by RFI resulted in useful data in the 110--200~MHz band and these 
calibrated spectra---without any Legendre polynomials subtracted---were used for foreground removal and signal detection.

Long duration laboratory tests of the receiver were done with the
antenna replaced by a variety of terminations: open, short and
impedance matched terminations and a resistor--inductor--capacitor network with $\Gamma_c(\nu)$ similar to that of
the SARAS~2 antenna.  All of these, on processing as above and fitted using a single smooth function as defined in \citet[]{2015ApJ...810....3S}, yielded residuals consistent with expected thermal noise.

\begin{figure}[htbp]
\begin{center}
\includegraphics[scale=0.3, width=\columnwidth]{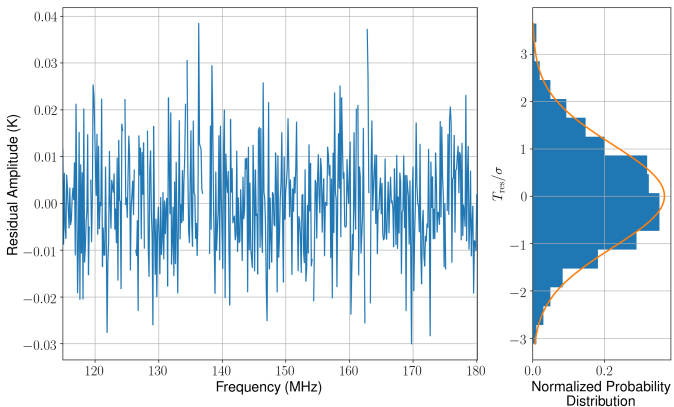}
\caption{Residuals obtained after fitting calibrated sky
  data, following RFI rejection, with a 7-th order polynomial model representing the
  foreground.  On the right is a histogram for the amplitudes along with the best-fit Gaussian.
  Since data rejection for RFI varies across channels, the channel amplitudes vary in their
  signal-to-noise ratio and, therefore, amplitudes are normalized by
  their 1$\sigma$ errors for the histogram.} 
\label{fig:res}
\end{center}
\end{figure}

The modeling of foreground in the sky data was performed by fitting polynomials. This also inevitably results in partial filtering out of the EoR signal. We adopt the global 21-cm templates predicted by the semi-numerical simulations of \citet{2016arXiv160902312C} as representative of currently allowed signals. Since these different EoR templates have different variations with frequency, we separately optimize for different templates the order of polynomial and frequency sub-band for their analysis to maximise the signal-to-noise ratio in the residual. This yields a set of residuals, individually optimized for the detection of different templates, and these are used below for deriving constraints on the EoR.  These residuals have root-mean-square (rms) noise of about 11~mK and a representative residual is shown in Fig.~\ref{fig:res}.

\section{Constraints on baryon thermal evolution}

Any EoR signal appearing in each of the residual spectra obtained after fitting data with appropriate polynomials would be attenuated
by the total efficiency $\eta_t$, shown in Fig.~\ref{fig:Ant2}, and ``high-pass filtered" due to the subtraction of the fitted polynomial from the data.
Corresponding to any plausible EoR signal we may thus construct a
``processed" EoR signal that is expected in the residual by fitting out a polynomial of the same order to the attenuated template.  We have confirmed via simulations that this polynomial fitting process is linear.

To test for the presence of any plausible EoR signal in the data residual, we compute the ratio of the likelihood of the residual containing the processed signal plus expected Gaussian noise
(the alternate hypothesis $H_1$), and the likelihood of the residual containing just noise (the null hypothesis $H_0$).  We assume both cases to be equally likely and hence assign uniform priors.  The
likelihoods are defined to be:

\begin{align}
P(D|M)= \prod_{i=1}^{N} \frac{1}{\sqrt{(2 \pi \sigma_i^2)}} e^{\frac{-(y_i-M_i)^2}{2\sigma_i^2}}, 
\end{align}
where $y_i$ is the data residual in the $i^{th}$ frequency channel,
$\sigma_i$ is the associated error, $M_i$ is the model amplitude at
that channel and $N$ is the number of independent frequency channels.
We derive the measurement noise $\sigma_i$ by accounting for all of
the data rejection for RFI, measurements of the system temperature,
absolute calibration of SARAS~2 and finally from differences between
adjacent channel data.  The likelihood ratio
\begin{equation}
\textrm{LR} = \prod_{i=1}^{N}  \frac{e^{\frac{-(y_i-M_i)^2}{2\sigma_i^2}}}{e^{\frac{-y_i^2}{2\sigma_i^2}}}
\end{equation}
is the ratio of likelihoods of
$M$ being the processed signal to that for $M$ being zero.

To determine the significance of the likelihood ratio corresponding
to any particular EoR signal template, we generate mock datasets with the
same $\sigma_i$ distribution as that in the data residual. One
dataset $D_1$ contains the processed EoR template plus noise, while
the second dataset $D_0$ contains only noise.  We compute likelihood ratios for $D_0$ and $D_1$ for multiple realizations of noise to derive the expected distributions of these likelihood ratios. These distributions are then used to infer the 
probabilities for false positives and false negatives for the likelihood ratio derived from the data depending on whether the ratio for  any EoR template exceeds unity or is
below unity \cite[Chapter 3]{9780135041352}.

\begin{figure}[htbp]
\centering
   \begin{subfigure}[]{}
   \includegraphics[width=1\linewidth]{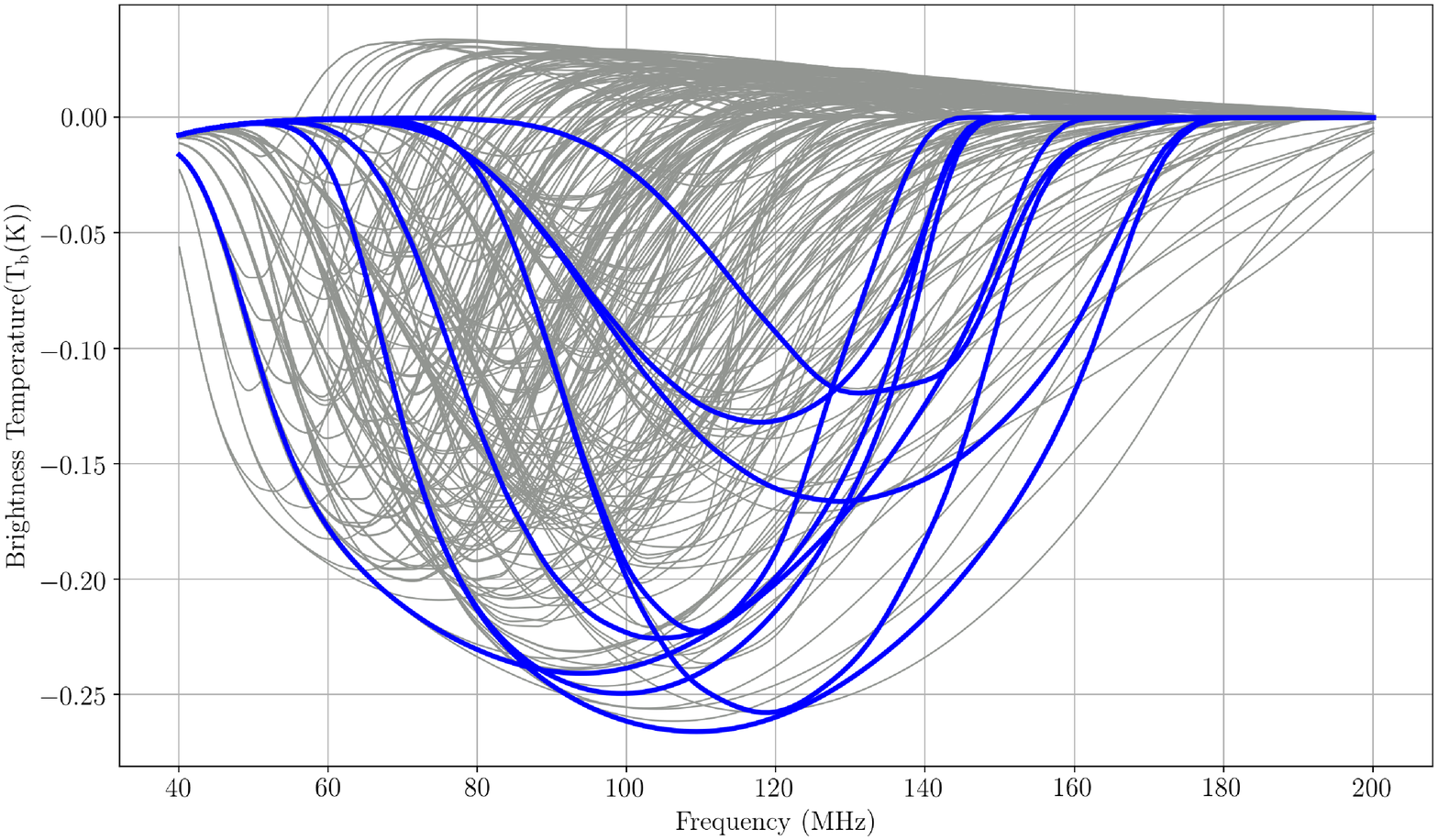}
   \label{fig:Ng1} 
\end{subfigure}
\begin{subfigure}[]{}
   \includegraphics[width=1\linewidth]{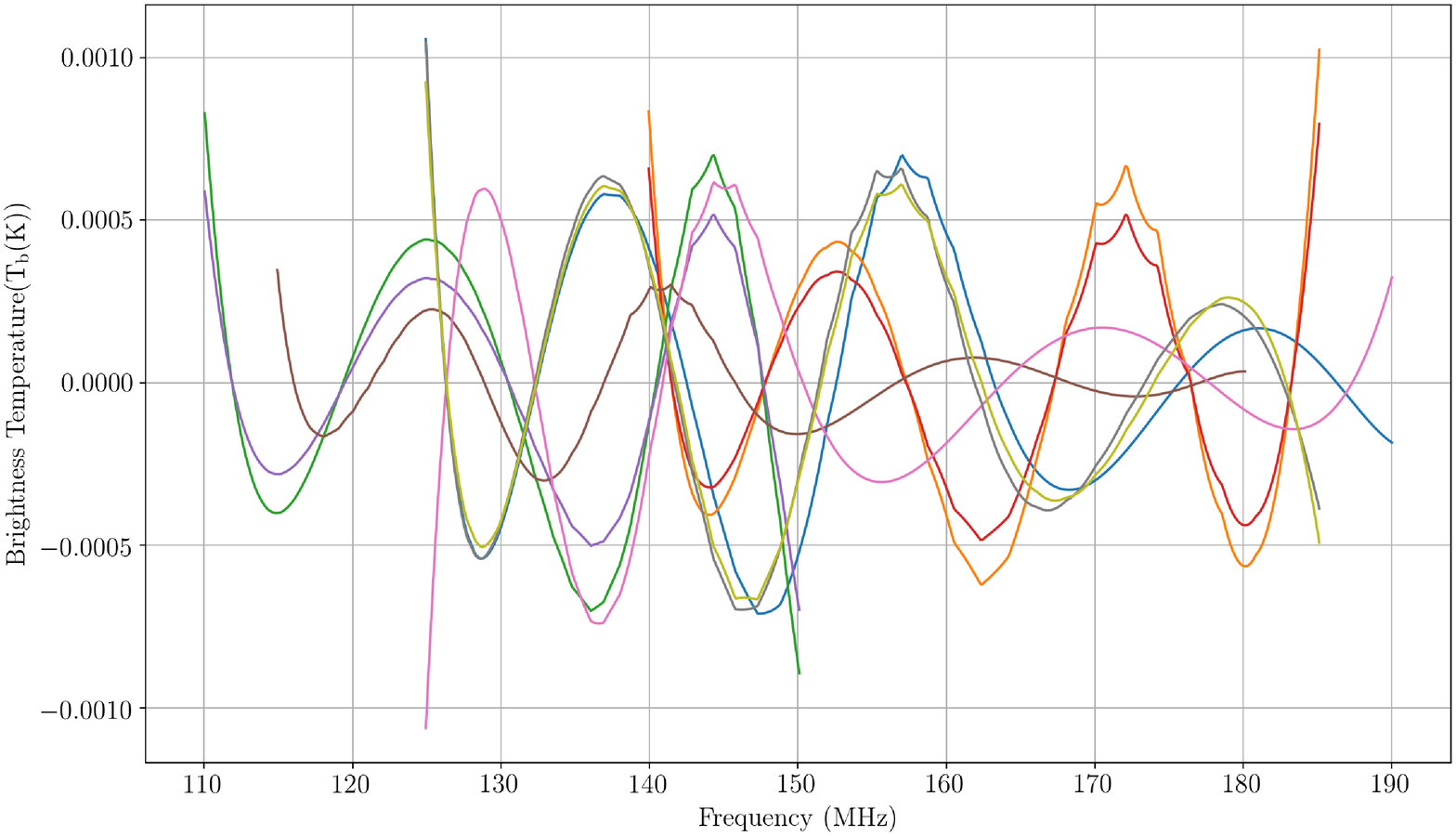}
   \label{fig:Ng2}
\end{subfigure}
\caption{Panel (a) shows the atlas of 21-cm templates highlighting those 9 that belong to the selected class of late heating and rapid reionization.  The grey curves show models that are not significantly constrained by the data. Panel (b) shows the processed EoR signals, which were obtained from the templates after attenuation by the antenna efficiency followed by high-pass filtering resulting from polynomial fits.}
\label{fig:temp1}
\end{figure}

Given the rms noise in the data and the amplitude of the processed signal, we infer that the data is sensitive to the class of signals corresponding to late heating or poor X-ray efficiency, with $f_X \leq 0.1$ (see \citet{2016arXiv160902312C} for details), along with peak $\frac{dT_b}{dz} \geq 120~\textrm{mK per unit redshift interval}$ corresponding to a rapid rate of reionization. 
We compute likelihood ratios from the residual data 
for the 21-cm templates that satisfied these criteria; there were 9 such cases out of the total of 264 in the atlas.  In Fig.~\ref{fig:temp1} we show these templates as well as their processed residuals.

\begin{figure}[htbp]
\begin{center}
\includegraphics[scale=0.3, width=\columnwidth]{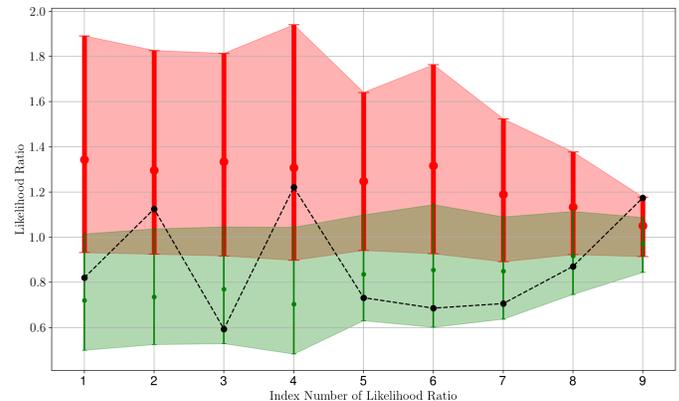}
\caption{Likelihood ratios for the 9 EoR templates that belong to the class defined in the text. For each template, we show the inferred likelihood ratio from the data (marked in black and connected using a dotted line).  We also show the extents (from $32^{\rm nd}$ to $68^{\rm th}$ percentile) of the distributions of $D_1$ and $D_0$ as shaded regions. The regions corresponding to the hypotheses $H_1$ (upper vertical column) and $H_0$ (lower vertical column) are shown in red and green respectively; their medians are shown using filled circles.}
\label{fig:temp2}
\end{center}
\end{figure}

We show in Fig.~\ref{fig:temp2} the likelihood ratios inferred from the data along with the
expected distributions of these ratios. For almost all of the signals belonging to this class the distributions of $D_1$ and $D_0$ are significantly separated and hence the data has the sensitivity to discriminate between the hypotheses $H_1$ and $H_0$ (presence or absence of the signal).  Of these allowed signals, six are disfavored in that their likelihood ratios place them in the domain of $H_0$, within its $32^{\rm nd}$ to $68^{\rm th}$ percentile band, and the probability of their being false negatives is in the range 14 \% to 28\%. Two signals have likelihood ratios within the $32^{\rm nd}$ to $68^{\rm th}$ percentile band of $H_1$; however, the probability that these are false alarms is as much as 25 to 30\%.  In the case of one signal---the one with index number 9 in the Figure---the data analysis leads to a result of relatively poorer significance. The class as a whole, taking into account all 9 signals, has likelihood ratios with an average probability of 31\% of being false negatives; therefore, the class of signals is more likely to be from $D_0$ than $D_1$. This implies that the data is more consistent with noise-only hypothesis as against the hypothesis in which noise and  template are present.  We thus disfavor this class of models with $f_X \leq 0.1$ and peak $\frac{dT_b}{dz} \geq 120~\textrm{mK per unit redshift interval}$ with $69\%$ confidence.

The models that are disfavored by SARAS~2 all lie in the area of
parameter space corresponding to late heating
\cite[]{2014Natur.506..197F}, actually a regime we might call very
late heating, in which cosmic reionization ends without the global
21-cm signal having reached emission. 
More specifically, SARAS~2 disfavors models that have late (i.e., weak)
X-ray heating and a rapid end to reionization (due, for example, to
large galaxies dominating star formation and a large mean free path
available within the ionized bubbles).

In summary, we disfavor the class of global 21-cm models that represent late heating or poor X-ray efficiency, with $f_X \leq 0.1$, and with peak $\frac{dT_b}{dz} \geq 120~\textrm{mK per unit redshift interval}$ corresponding to a rapid rate of reionization with $69\%$ confidence.  
These results are intriguing and we are devising better strategies for foreground modeling towards minimising loss in signal amplitudes.

\section*{Acknowledgement}

We thank the anonymous referee for their valuable comments that led to 
improved methods of data analysis.  We thank staff at Gauribidanur Field Station for assistance with
system tests and measurements, and the Mechanical 
and Electronics Engineering Groups at Raman Research Institute for
building and assembling SARAS~2. Santosh Harish and Divya Jayasankar
implemented real-time software and monitoring. Logistics and technical
support for observations was provided by Indian Astronomical
Observatory, Leh operated by Indian Institute of Astrophysics, and
Timbaktu Collective, India. For R.B.\ and A.C.\, this
project/publication was made possible through the support of a grant
from the John Templeton Foundation. The opinions expressed in this
publication are those of the author(s) and do not necessarily reflect
the views of the John Templeton Foundation.


\begin{thebibliography}{}
\expandafter\ifx\csname natexlab\endcsname\relax\def\natexlab#1{#1}\fi

\bibitem[{{Barkana}(2016)}]{2016PhR...645....1B}
{Barkana}, R. 2016, \physrep, 645, 1

\bibitem[{{Barkana} \& {Loeb}(2001)}]{2001PhR...349..125B}
{Barkana}, R., \& {Loeb}, A. 2001, \physrep, 349, 125

\bibitem[{{Beardsley} {et~al.}(2016){Beardsley}, {Hazelton}, {Sullivan},
  {Carroll}, {Barry}, {Rahimi}, {Pindor}, {Trott}, {Line}, {Jacobs}, {Morales},
  {Pober}, {Bernardi}, {Bowman}, {Busch}, {Briggs}, {Cappallo}, {Corey}, {de
  Oliveira-Costa}, {Dillon}, {Emrich}, {Ewall-Wice}, {Feng}, {Gaensler},
  {Goeke}, {Greenhill}, {Hewitt}, {Hurley-Walker}, {Johnston-Hollitt},
  {Kaplan}, {Kasper}, {Kim}, {Kratzenberg}, {Lenc}, {Loeb}, {Lonsdale},
  {Lynch}, {McKinley}, {McWhirter}, {Mitchell}, {Morgan}, {Neben},
  {Thyagarajan}, {Oberoi}, {Offringa}, {Ord}, {Paul}, {Prabu}, {Procopio},
  {Riding}, {Rogers}, {Roshi}, {Udaya Shankar}, {Sethi}, {Srivani},
  {Subrahmanyan}, {Tegmark}, {Tingay}, {Waterson}, {Wayth}, {Webster},
  {Whitney}, {Williams}, {Williams}, {Wu}, \& {Wyithe}}]{2016ApJ...833..102B}
{Beardsley}, A.~P., {Hazelton}, B.~J., {Sullivan}, I.~S., {et~al.} 2016, \apj,
  833, 102

\bibitem[{{Bernardi} {et~al.}(2015){Bernardi}, {McQuinn}, \&
  {Greenhill}}]{2015ApJ...799...90B}
{Bernardi}, G., {McQuinn}, M., \& {Greenhill}, L.~J. 2015, \apj, 799, 90

\bibitem[{{Bernardi} {et~al.}(2016){Bernardi}, {Zwart}, {Price}, {Greenhill},
  {Mesinger}, {Dowell}, {Eftekhari}, {Ellingson}, {Kocz}, \&
  {Schinzel}}]{2016MNRAS.461.2847B}
{Bernardi}, G., {Zwart}, J.~T.~L., {Price}, D., {et~al.} 2016, \mnras, 461,
  2847

\bibitem[{{Bowman} \& {Rogers}(2010)}]{2010Natur.468..796B}
{Bowman}, J.~D., \& {Rogers}, A.~E.~E. 2010, \nat, 468, 796

\bibitem[{{Ciardi} \& {Ferrara}(2005)}]{2005SSRv..116..625C}
{Ciardi}, B., \& {Ferrara}, A. 2005, \ssr, 116, 625

\bibitem[{{Cohen} {et~al.}(2016){Cohen}, {Fialkov}, {Barkana}, \&
  {Lotem}}]{2016arXiv160902312C}
{Cohen}, A., {Fialkov}, A., {Barkana}, R., \& {Lotem}, M. 2016, ArXiv e-prints,
  arXiv:1609.02312

\bibitem[{{Fan} {et~al.}(2006){Fan}, {Carilli}, \&
  {Keating}}]{2006ARA&A..44..415F}
{Fan}, X., {Carilli}, C.~L., \& {Keating}, B. 2006, \araa, 44, 415

\bibitem[{{Fialkov} {et~al.}(2014){Fialkov}, {Barkana}, \&
  {Visbal}}]{2014Natur.506..197F}
{Fialkov}, A., {Barkana}, R., \& {Visbal}, E. 2014, \nat, 506, 197

\bibitem[{{Field}(1958)}]{Field:1958}
{Field}, G.~B. 1958, Proceedings of the IRE, 46, 240

\bibitem[{{Furlanetto}(2006)}]{2006MNRAS.371..867F}
{Furlanetto}, S.~R. 2006, \mnras, 371, 867

\bibitem[{{Furlanetto} {et~al.}(2006){Furlanetto}, {Oh}, \&
  {Briggs}}]{2006PhR...433..181F}
{Furlanetto}, S.~R., {Oh}, S.~P., \& {Briggs}, F.~H. 2006, \physrep, 433, 181

\bibitem[{Haiman(2016)}]{Haiman2016}
Haiman, Z. 2016, Cosmic Reionization and the First Nonlinear Structures in the
  Universe (Cham: Springer International Publishing), 1--22

\bibitem[{Hampel(1974)}]{10.2307/2285666}
Hampel, F.~R. 1974, Journal of the American Statistical Association, 69, 383

\bibitem[{{Harker}(2015)}]{2015MNRAS.449L..21H}
{Harker}, G.~J.~A. 2015, \mnras, 449, L21

\bibitem[{{Harker} {et~al.}(2016){Harker}, {Mirocha}, {Burns}, \&
  {Pritchard}}]{2016MNRAS.455.3829H}
{Harker}, G.~J.~A., {Mirocha}, J., {Burns}, J.~O., \& {Pritchard}, J.~R. 2016,
  \mnras, 455, 3829

\bibitem[{Hinshaw {et~al.}(2013)Hinshaw, Larson, Komatsu, Spergel, Bennett,
  Dunkley, Nolta, Halpern, Hill, Odegard, Page, Smith, Weiland, Gold, Jarosik,
  Kogut, Limon, Meyer, Tucker, Wollack, \& Wright}]{0067-0049-208-2-19}
Hinshaw, G., Larson, D., Komatsu, E., {et~al.} 2013, The Astrophysical Journal
  Supplement Series, 208, 19

\bibitem[{Kay(1998)}]{9780135041352}
Kay, S.~M. 1998, Fundamentals of Statistical Signal Processing, Volume II:
  Detection Theory (Prentice Hall)

\bibitem[{{Liu} {et~al.}(2013){Liu}, {Pritchard}, {Tegmark}, \&
  {Loeb}}]{2013PhRvD..87d3002L}
{Liu}, A., {Pritchard}, J.~R., {Tegmark}, M., \& {Loeb}, A. 2013, \prd, 87,
  043002

\bibitem[{{Liu} \& {Tegmark}(2012)}]{2012MNRAS.419.3491L}
{Liu}, A., \& {Tegmark}, M. 2012, \mnras, 419, 3491

\bibitem[{{Madau} \& {Fragos}(2016)}]{2016arXiv160607887M}
{Madau}, P., \& {Fragos}, T. 2016, ArXiv e-prints, arXiv:1606.07887

\bibitem[{{Madau} {et~al.}(1997){Madau}, {Meiksin}, \& {Rees}}]{MMR}
{Madau}, P., {Meiksin}, A., \& {Rees}, M.~J. 1997, \apj, 475, 429

\bibitem[{{McGreer} {et~al.}(2015){McGreer}, {Mesinger}, \&
  {D'Odorico}}]{2015MNRAS.447..499M}
{McGreer}, I.~D., {Mesinger}, A., \& {D'Odorico}, V. 2015, \mnras, 447, 499

\bibitem[{{McQuinn}(2016)}]{2016ARA&A..54..313M}
{McQuinn}, M. 2016, \araa, 54, 313

\bibitem[{Meys(1978)}]{meys1978wave}
Meys, R. 1978, IEEE Transactions on Microwave Theory and Techniques, 26, 34

\bibitem[{{Mirocha} {et~al.}(2017){Mirocha}, {Furlanetto}, \&
  {Sun}}]{2017MNRAS.464.1365M}
{Mirocha}, J., {Furlanetto}, S.~R., \& {Sun}, G. 2017, \mnras, 464, 1365

\bibitem[{{Mirocha} {et~al.}(2013){Mirocha}, {Harker}, \&
  {Burns}}]{2013ApJ...777..118M}
{Mirocha}, J., {Harker}, G.~J.~A., \& {Burns}, J.~O. 2013, \apj, 777, 118

\bibitem[{{Monsalve} {et~al.}(2017){Monsalve}, {Rogers}, {Bowman}, \&
  {Mozdzen}}]{2017ApJ...835...49M}
{Monsalve}, R.~A., {Rogers}, A.~E.~E., {Bowman}, J.~D., \& {Mozdzen}, T.~J.
  2017, \apj, 835, 49

\bibitem[{{Morandi} \& {Barkana}(2012)}]{2012MNRAS.424.2551M}
{Morandi}, A., \& {Barkana}, R. 2012, \mnras, 424, 2551

\bibitem[{{Nuttall}(1981)}]{1981ITASS..29...84N}
{Nuttall}, A.~H. 1981, IEEE Transactions on Acoustics Speech and Signal
  Processing, 29, 84

\bibitem[{{Patil} {et~al.}(2017){Patil}, {Yatawatta}, {Koopmans}, {de Bruyn},
  {Brentjens}, {Zaroubi}, {Asad}, {Hatef}, {Jelic}, {Mevius}, {Offringa},
  {Pandey}, {Vedantham}, {Abdalla}, {Brouw}, {Chapman}, {Ciardi}, {Gehlot},
  {Ghosh}, {Harker}, {Iliev}, {Kakiichi}, {Majumdar}, {Silva}, {Mellema},
  {Schaye}, {Vrbanec}, \& {Wijnholds}}]{2017arXiv170208679P}
{Patil}, A.~H., {Yatawatta}, S., {Koopmans}, L.~V.~E., {et~al.} 2017, ArXiv
  e-prints, arXiv:1702.08679

\bibitem[{{Patra} {et~al.}(2013){Patra}, {Subrahmanyan}, {Raghunathan}, \&
  {Udaya Shankar}}]{2013ExA....36..319P}
{Patra}, N., {Subrahmanyan}, R., {Raghunathan}, A., \& {Udaya Shankar}, N.
  2013, Experimental Astronomy, 36, 319

\bibitem[{{Planck Collaboration} {et~al.}(2016){Planck Collaboration}, {Adam},
  {Aghanim}, {Ashdown}, {Aumont}, {Baccigalupi}, {Ballardini}, {Banday},
  {Barreiro}, {Bartolo}, {Basak}, {Battye}, {Benabed}, {Bernard}, {Bersanelli},
  {Bielewicz}, {Bock}, {Bonaldi}, {Bonavera}, {Bond}, {Borrill}, {Bouchet},
  {Boulanger}, {Bucher}, {Burigana}, {Calabrese}, {Cardoso}, {Carron},
  {Chiang}, {Colombo}, {Combet}, {Comis}, {Couchot}, {Coulais}, {Crill},
  {Curto}, {Cuttaia}, {Davis}, {de Bernardis}, {de Rosa}, {de Zotti},
  {Delabrouille}, {Di Valentino}, {Dickinson}, {Diego}, {Dor{\'e}}, {Douspis},
  {Ducout}, {Dupac}, {Elsner}, {En{\ss}lin}, {Eriksen}, {Falgarone}, {Fantaye},
  {Finelli}, {Forastieri}, {Frailis}, {Fraisse}, {Franceschi}, {Frolov},
  {Galeotta}, {Galli}, {Ganga}, {G{\'e}nova-Santos}, {Gerbino}, {Ghosh},
  {Gonz{\'a}lez-Nuevo}, {G{\'o}rski}, {Gruppuso}, {Gudmundsson}, {Hansen},
  {Helou}, {Henrot-Versill{\'e}}, {Herranz}, {Hivon}, {Huang}, {Ili{\'c}},
  {Jaffe}, {Jones}, {Keih{\"a}nen}, {Keskitalo}, {Kisner}, {Knox},
  {Krachmalnicoff}, {Kunz}, {Kurki-Suonio}, {Lagache}, {L{\"a}hteenm{\"a}ki},
  {Lamarre}, {Langer}, {Lasenby}, {Lattanzi}, {Lawrence}, {Le Jeune},
  {Levrier}, {Lewis}, {Liguori}, {Lilje}, {L{\'o}pez-Caniego}, {Ma},
  {Mac{\'{\i}}as-P{\'e}rez}, {Maggio}, {Mangilli}, {Maris}, {Martin},
  {Mart{\'{\i}}nez-Gonz{\'a}lez}, {Matarrese}, {Mauri}, {McEwen}, {Meinhold},
  {Melchiorri}, {Mennella}, {Migliaccio}, {Miville-Desch{\^e}nes}, {Molinari},
  {Moneti}, {Montier}, {Morgante}, {Moss}, {Naselsky}, {Natoli}, {Oxborrow},
  {Pagano}, {Paoletti}, {Partridge}, {Patanchon}, {Patrizii}, {Perdereau},
  {Perotto}, {Pettorino}, {Piacentini}, {Plaszczynski}, {Polastri}, {Polenta},
  {Puget}, {Rachen}, {Racine}, {Reinecke}, {Remazeilles}, {Renzi}, {Rocha},
  {Rossetti}, {Roudier}, {Rubi{\~n}o-Mart{\'{\i}}n}, {Ruiz-Granados},
  {Salvati}, {Sandri}, {Savelainen}, {Scott}, {Sirri}, {Sunyaev}, {Suur-Uski},
  {Tauber}, {Tenti}, {Toffolatti}, {Tomasi}, {Tristram}, {Trombetti},
  {Valiviita}, {Van Tent}, {Vielva}, {Villa}, {Vittorio}, {Wandelt}, {Wehus},
  {White}, {Zacchei}, \& {Zonca}}]{2016A&A...596A.108P}
{Planck Collaboration}, {Adam}, R., {Aghanim}, N., {et~al.} 2016, \aap, 596,
  A108

\bibitem[{{Pober} {et~al.}(2015){Pober}, {Ali}, {Parsons}, {McQuinn},
  {Aguirre}, {Bernardi}, {Bradley}, {Carilli}, {Cheng}, {DeBoer}, {Dexter},
  {Furlanetto}, {Grobbelaar}, {Horrell}, {Jacobs}, {Klima}, {Kohn}, {Liu},
  {MacMahon}, {Maree}, {Mesinger}, {Moore}, {Razavi-Ghods}, {Stefan},
  {Walbrugh}, {Walker}, \& {Zheng}}]{2015ApJ...809...62P}
{Pober}, J.~C., {Ali}, Z.~S., {Parsons}, A.~R., {et~al.} 2015, \apj, 809, 62

\bibitem[{{Pritchard} \& {Loeb}(2012)}]{2012RPPh...75h6901P}
{Pritchard}, J.~R., \& {Loeb}, A. 2012, Reports on Progress in Physics, 75,
  086901

\bibitem[{{Sathyanarayana Rao} {et~al.}(2015){Sathyanarayana Rao},
  {Subrahmanyan}, {Udaya Shankar}, \& {Chluba}}]{2015ApJ...810....3S}
{Sathyanarayana Rao}, M., {Subrahmanyan}, R., {Udaya Shankar}, N., \& {Chluba},
  J. 2015, \apj, 810, 3

\bibitem[{{Sathyanarayana Rao} {et~al.}(2017{\natexlab{a}}){Sathyanarayana
  Rao}, {Subrahmanyan}, {Udaya Shankar}, \& {Chluba}}]{2017AJ....153...26S}
---. 2017{\natexlab{a}}, \aj, 153, 26

\bibitem[{{Sathyanarayana Rao} {et~al.}(2017{\natexlab{b}}){Sathyanarayana
  Rao}, {Subrahmanyan}, {Udaya Shankar}, \& {Chluba}}]{2017ApJ...840...33S}
---. 2017{\natexlab{b}}, \apj, 840, 33

\bibitem[{{Sault} {et~al.}(1995){Sault}, {Teuben}, \&
  {Wright}}]{1995ASPC...77..433S}
{Sault}, R.~J., {Teuben}, P.~J., \& {Wright}, M.~C.~H. 1995, in Astronomical
  Society of the Pacific Conference Series, Vol.~77, Astronomical Data Analysis
  Software and Systems IV, ed. R.~A. {Shaw}, H.~E. {Payne}, \& J.~J.~E.
  {Hayes}, 433

\bibitem[{{Shaver} {et~al.}(1999){Shaver}, {Windhorst}, {Madau}, \& {de
  Bruyn}}]{1999A&A...345..380S}
{Shaver}, P.~A., {Windhorst}, R.~A., {Madau}, P., \& {de Bruyn}, A.~G. 1999,
  \aap, 345, 380

\bibitem[{{Sokolowski} {et~al.}(2015{\natexlab{a}}){Sokolowski}, {Tremblay},
  {Wayth}, {Tingay}, {Clarke}, {Roberts}, {Waterson}, {Ekers}, {Hall}, {Lewis},
  {Mossammaparast}, {Padhi}, {Schlagenhaufer}, {Sutinjo}, \&
  {Tickner}}]{2015PASA...32....4S}
{Sokolowski}, M., {Tremblay}, S.~E., {Wayth}, R.~B., {et~al.}
  2015{\natexlab{a}}, \pasa, 32, e004

\bibitem[{{Sokolowski} {et~al.}(2015{\natexlab{b}}){Sokolowski}, {Wayth},
  {Tremblay}, {Tingay}, {Waterson}, {Tickner}, {Emrich}, {Schlagenhaufer},
  {Kenney}, \& {Padhi}}]{2015ApJ...813...18S}
{Sokolowski}, M., {Wayth}, R.~B., {Tremblay}, S.~E., {et~al.}
  2015{\natexlab{b}}, \apj, 813, 18

\bibitem[{{Vedantham} {et~al.}(2014){Vedantham}, {Koopmans}, {de Bruyn},
  {Wijnholds}, {Ciardi}, \& {Brentjens}}]{2014MNRAS.437.1056V}
{Vedantham}, H.~K., {Koopmans}, L.~V.~E., {de Bruyn}, A.~G., {et~al.} 2014,
  \mnras, 437, 1056

\bibitem[{{Voytek} {et~al.}(2014){Voytek}, {Natarajan}, {J{\'a}uregui
  Garc{\'{\i}}a}, {Peterson}, \& {L{\'o}pez-Cruz}}]{2014ApJ...782L...9V}
{Voytek}, T.~C., {Natarajan}, A., {J{\'a}uregui Garc{\'{\i}}a}, J.~M.,
  {Peterson}, J.~B., \& {L{\'o}pez-Cruz}, O. 2014, \apjl, 782, L9

\bibitem[{{Wouthuysen}(1952)}]{Wouthuysen:1952}
{Wouthuysen}, S.~A. 1952, Astronomical Journal, 57, 31

\bibitem[{{Zaroubi}(2013)}]{2013ASSL..396...45Z}
{Zaroubi}, S. 2013, in Astrophysics and Space Science Library, Vol. 396, The
  First Galaxies, ed. T.~{Wiklind}, B.~{Mobasher}, \& V.~{Bromm}, 45

\bibitem[{{Zheng} {et~al.}(2017){Zheng}, {Wang}, {Rhoads}, {Infante},
  {Malhotra}, {Hu}, {Walker}, {Jiang}, {Jiang}, {Hibon}, {Gonzalez}, {Kong},
  {Zheng}, {Galaz}, \& {Barrientos}}]{2017arXiv170302985Z}
{Zheng}, Z.-Y., {Wang}, J., {Rhoads}, J., {et~al.} 2017, ArXiv e-prints,
  arXiv:1703.02985

\end{thebibliography}
\end{document}